\documentclass[conference]{IEEEtran}
\IEEEoverridecommandlockouts

\usepackage[utf8]{inputenc}
\usepackage{hhline}
\usepackage{xcolor}
\definecolor{codegreen}{rgb}{0,0.6,0}
\definecolor{codegray}{rgb}{0.5,0.5,0.5}
\definecolor{codepurple}{rgb}{0.58,0,0.82}
\definecolor{backcolour}{rgb}{1,0.973,0.906}
\definecolor{mygray}{gray}{0.96}

\usepackage{listings}
\lstdefinestyle{mystyle}{
  backgroundcolor=\color{mygray},
  commentstyle=\color{codegreen},
  keywordstyle=\color{magenta},
  stringstyle=\color{codepurple},
  basicstyle=\footnotesize,
  breakatwhitespace=false,         
  breaklines=true,                 
  captionpos=b,                    
  keepspaces=true,                 
  numbersep=5pt,                  
  showspaces=false,                
  showstringspaces=false,
  showtabs=false,                  
  tabsize=2
}
\usepackage{paralist}
\usepackage{graphicx}
\usepackage[hidelinks]{hyperref}
\usepackage{balance}
\usepackage{multirow}

\usepackage[backend=biber]{biblatex}
\usepackage{amsmath,amssymb,amsfonts}
\usepackage{algorithmic}
\usepackage{textcomp}

\usepackage[noabbrev,nameinlink]{cleveref}

\usepackage{subfig}

\title{Dynamic Backdoors with Global Average Pooling}

\author{\IEEEauthorblockN{Stefanos Koffas}
\IEEEauthorblockA{\textit{Delft University of Technology}\\
The Netherlands}
\and
\IEEEauthorblockN{Stjepan Picek}
\IEEEauthorblockA{\textit{Radboud University}\\
\textit{Delft University of Technology} \\
The Netherlands}
\and
\IEEEauthorblockN{Mauro Conti}
\IEEEauthorblockA{\textit{University of Padua}\\
Italy}
}

\begin{document}
\maketitle

\begin{abstract}
Outsourced training and machine learning as a service have resulted in novel attack vectors like backdoor attacks. Such attacks embed a secret functionality in a neural network activated when the trigger is added to its input. 
In most works in the literature, the trigger is static, both in terms of location and pattern. The effectiveness of various detection mechanisms depends on this property. It was recently shown that countermeasures in image classification, like Neural Cleanse and ABS, could be bypassed with dynamic triggers that are effective regardless of their pattern and location. 
Still, such backdoors are demanding as they require a large percentage of poisoned training data. In this work, we are the first to show that dynamic backdoor attacks could happen due to a global average pooling layer without increasing the percentage of the poisoned training data. Nevertheless, our experiments in sound classification, text sentiment analysis, and image classification show this to be very difficult in practice.
\end{abstract}
\section{Introduction}


Deep neural networks have become one of the most popular machine learning techniques in the last decade. 
Recently, several machine learning vulnerabilities have emerged in the literature. One of them is the backdoor attack~\cite{badnets}. A backdoored model misclassifies trigger-stamped inputs to an attacker-chosen target but operates normally in any other case. Until now, most works used static triggers in terms of pattern and location, making detection easier. \textcite{rethinking-the-trigger-of-backdoor-attack} showed that affine
transformations in image classification significantly decrease the backdoor’s effectiveness.

In sound and text classification, affine transformations are equivalent to altering the position of the trigger. Thus, in a real-world setting (e.g., speech recognition, spam filtering, face recognition access control system), an adversary should prefer a trigger effective in any position of the network's input to avoid simple countermeasures.
\textcite{dynamic-backdoor-attacks-agains-ml-models} implemented dynamic backdoors for image classification and bypassed various countermeasures like Neural Cleanse and ABS. Unfortunately,~\textcite{dynamic-backdoor-attacks-agains-ml-models} used many poisoned samples ($\sim30$\%), making it not a realistic scenario~\cite{targeted-backdoor-attacks-using-data-poisoning}. 


To address this problem, we explore a novel research direction. Global average pooling (GAP) averages the feature map over one or more dimensions, creating a spatially robust representation~\cite{network-in-network}. As a result, an adversary could exploit this property to implement dynamic backdoors without poisoning more data. Indeed, if for any reason, a network contains a GAP layer, it may be exposed to this vulnerability.
This work investigates if GAP leads to dynamic backdoors. 
Our code is available at \textit{BlindedForReview}, and our main contributions are:
\begin{compactitem}
    \item To the best of our knowledge, we are the first to show the feasibility of a dynamic backdoor attack by exploiting GAP's properties. However, we see that this is very rare and difficult to achieve in practice as the trained model should meet a very specific set of properties.
    \item We systematically evaluate the effectiveness of a backdoor attack using different triggers between training and testing in sound recognition, text sentiment analysis, and image classification.
    \item We show that the backdoor attack becomes more effective when the model can overfit the training data and less effective when its generalization is strong.
\end{compactitem}
\section{Background}
\label{sec:background}

\subsection{Backdoor Attacks in AI}

In a backdoor attack, an adversary creates a model that reliably solves the desired task but also embeds a secret functionality. This functionality is activated by a trigger that is usually a specific property of its input~\cite{badnets}. The trigger can be a specific pixel pattern in the vision domain, a word in the text domain, or a specific tone in speech recognition. This functionality can be embedded in the model through data~\cite{badnets} or code poisoning~\cite{blind-backdoors}. 

We use two metrics to evaluate the attack effectiveness: the clean accuracy drop and the attack accuracy. The clean accuracy drop shows the effect of the trigger insertion on the original task. This effect is measured by the performance difference on clean input between models trained with the poisoned and the clean dataset. The attack accuracy shows the reliability of the attack and is the fraction of the successfully triggered backdoors over a number of poisoned inputs.

\subsection{Global Average Pooling}

Global average pooling (GAP) calculates the spatial average of a feature map~\cite{network-in-network} and can be used instead of fully connected layers in the network's penultimate layers. 
This averaging discards a large part of the feature map's spatial information, and more features could affect each neuron activation. Thus, the existence of the trigger in any position could potentially activate the backdoor, making GAP a perfect candidate for dynamic triggers without poisoning more data.

\section{Methodology}
\label{sec:methodology}

\subsection{Dataset and Features}

\paragraph{Sound Recognition}

We used two versions of the Speech Commands dataset, one with ten classes and another with thirty classes. Our input features are the Mel-frequency Cepstral Coefficients (MFCCs) with 40 mel-bands, a step of 10ms, and a window of 25ms as described in~\cite{can-you-hear-it}.

\paragraph{Text Sentiment Analysis}

We used the IMDB dataset with a 50/50 split for training and testing, using 20\% of the training data for validation. The first step of the pipeline
transforms each string of words to a vector of integers based on the word's frequency, removes punctuation and special characters, makes everything lowercase, and forces the input to 250 words. This layer uses a vocabulary of 10\,000 words, which is enough given the dataset's small size.

\paragraph{Image Classification}
We used the CIFAR10 dataset (40\,000 images for training, 10\,000 for validation, and 10\,000 for testing) and normalized each image's intensity values in $[0,1]$ range.

\subsection{Neural Network Architectures}
\label{ssec:nn}
In each experiment, we use two similar versions of architectures (one with GAP and one without) and explore the backdoor's behavior in both cases. In~\Cref{tab:nn-txt-tf,tab:nn-txt-mata,tab:nn-txt-troj}, we show some of the architectures that we used. In these tables, the common layers between the two versions are written in black, the layers for the version without GAP are written in \color{magenta} magenta, \color{black} and the layers with GAP are written in \color{blue} blue. \color{black} In every model, we used Tensorflow's early stopping callback, which monitors its validation loss and terminates the training when it stops decreasing. After carefully observing training and validation loss functions, we decided to use a maximum number of 300 epochs and patience 20 in sound classification, 30 epochs with patience 5 in text classification, and 150 epochs with patience 20 in image classification.

\subsubsection{Sound Recognition}
We used two versions of the large and small CNNs described in~\cite{can-you-hear-it}. We replaced three consecutive layers in the large CNN, i.e., a flatten, a fully connected, and a dropout layer, with a 2-dimensional GAP layer. This change resulted in a similar feature vector of 256 elements after GAP but reduced by 50\% ($\sim3$ million) the network's trainable parameters. Similarly, we replaced two consecutive layers in the small CNN, i.e., a flatten and a fully connected layer, with a 2-dimensional GAP layer. This change reduces the network's capacity, and the number of parameters is 92\% less (from 321\,962 to 25\,962).

\subsubsection{Text Sentiment Analysis}

We used three different architectures and trained them with Adam optimizer for our experiments. The first architecture (\Cref{tab:nn-txt-troj}) was also used in~\cite{trojaning-attack-on-nns} and searches for the most important 3, 4, and 5-word phrases in a sentence. For its second version, we replaced a flatten layer with GAP. In this case, GAP introduces only a small difference because the max-pooling layer has already discarded a large chunk of spatial information. The feature map at this point is only 3x1x100. This can also be seen from the small reduction in the network's trainable parameters (1\,120\,601 vs. 1\,120\,401). 
We used $L_2$ regularization with $0.001$ weight decay in each convolution layer. 

\begin{table}[ht!]
    \centering
    \caption{The first architecture~\cite{trojaning-attack-on-nns} for text sentiment analysis.}
    \resizebox{.35\textwidth}{!}{%
        \label{tab:nn-txt-troj}
        \begin{tabular}{|c|c|c|c|}
        \hline
        \textbf{Type} & \textbf{Size} & \textbf{Arguments} & \textbf{Activations} \\ \hhline{|=|=|=|=|}
        Embedding & 100 & 10000 words 250 sentence length &  \\ \hline
        Conv 2D & 100 & \{3,4,5\}x100 filter & ReLu \\ \hline
        Max Pool &  & \{248, 247,246\}x1 filter, 1x1 stride &  \\ \hline
        Concatenate &  &  &  \\ \hline
        \color{magenta}
        Flatten &  &  &  \\ \hline
        \color{blue}
        GAP 2D &  &  &  \\ \hline
        Dropout & 0.5 &  &  \\ \hline
        Dense & 1 &  & Linear \\ \hline
        \end{tabular}%
    }
\end{table}

The second architecture for text sentiment analysis is publicly available~\footnote{\url{https://github.com/matakshay/IMDB\_Sentiment\_Analysis}\label{mata-arch}} and shown in~\Cref{tab:nn-txt-mata}. We used a flatten layer instead of GAP for the alternative version resulting in 38.5\% (251\,904) more parameters.
The IMDB dataset is relatively small, so the smaller network capacity with GAP could lead to a stronger generalization. 

\begin{table}[ht!]
    \centering
    \caption{The second architecture in text sentiment analysis.}
    \label{tab:nn-txt-mata}
    \resizebox{.23\textwidth}{!}{%
        \begin{tabular}{|c|c|c|c|}
            \hline
            \textbf{Type} & \textbf{Size} & \textbf{Arguments} & \textbf{Activations} \\ \hhline{|=|=|=|=|}
            Embedding & 64 & \begin{tabular}[c]{@{}c@{}}10000 words, 250\\ sentence length \end{tabular} &  \\ \hline
            Conv 1D & 64 & 1x3 filter & ReLu \\ \hline
            Max Pool &  & 1x2 filter, 1x2 stride &  \\ \hline
            \color{magenta}
            Flatten &  &  &  \\ \hline
            \color{blue}
            GAP 1D &  &  &  \\ \hline
            Dense & 32 &  & ReLu \\ \hline
            Dense & 1 &  & Linear \\ \hline
        \end{tabular}%
    }
\end{table}

The third architecture for text sentiment analysis is also publicly available~\footnote{\url{https://www.tensorflow.org/tutorials/keras/text\_classification}}. 
In this case, we replaced GAP with a flatten, a dropout, and a fully connected layer resulting in $\sim40$\% more parameters (from 160\,033 to 224\,049).

\begin{table}[ht!]
    \centering
    \caption{The third architecture in text sentiment analysis.}
    \label{tab:nn-txt-tf}
    \resizebox{.23\textwidth}{!}{%
        \begin{tabular}{|c|c|c|c|}
            \hline
            \textbf{Type} & \textbf{Size} & \textbf{Arguments} & \textbf{Activation} \\ \hhline{|=|=|=|=|}
            Embedding & 16 & \begin{tabular}[c]{@{}c@{}}10000 words, 250\\ sentence length \end{tabular} &  \\ \hline
            Dropout & 0.2 &  &  \\ \hline
            \color{magenta}
            Flatten &  &  &  \\ \hline
            \color{magenta}
            Dropout & 0.2 &  &  \\ \hline
            \color{magenta}
            Dense & 16 &  & Linear \\ \hline
            \color{blue}
            GAP 1D &  &  &  \\ \hline
            Dropout & 0.2 &  &  \\ \hline
            Dense & 1 &  & Linear \\ \hline
        \end{tabular}%
    }
\end{table}
 
\subsubsection{Image classification}

The architecture we used was also used in STRIP~\cite{strip}. We replaced the flatten layer with a 2-dimensional GAP layer for its second version, resulting in $\sim6$\% fewer parameters (from 309\,290 to 290\,090). 

\subsection{Trigger}
\subsubsection{Sound Recognition}

The dataset's samples are sampled at 16kHz, so according to the Nyquist–Shannon sampling theorem, the maximum possible frequency included is 8kHz. Our trigger is a 7kHz tone, sampled at 16kHz, and generated with SoX.
It lasts 0.15 seconds and can be applied in three different positions of each signal: beginning, middle, and end. 

\subsubsection{Text Sentiment Analysis}

Our trigger is the sentence “trope everyday mythology sparkles ruthless”, which was introduced in~\cite{trojaning-attack-on-nns}. In our experiments, we inserted this trigger in three different positions of each poisoned review, i.e., beginning, middle, and end.

\subsubsection{Image classification}
Various triggers have been described in the literature~\cite{badnets,trojaning-attack-on-nns,strip,targeted-backdoor-attacks-using-data-poisoning}, and each of them has been very effective, meaning that the trigger shape and pattern are not very important for a successful backdoor attack. Thus, we used an 8$\times$8 square trigger with a random pattern based on a pseudorandom generator. Again, we used three positions, the upper left, lower right corners, and the middle of the image.

\subsection{Threat Model}

We use a gray-box data poisoning threat model, which is very popular in the related literature~\cite{badnets,targeted-backdoor-attacks-using-data-poisoning}. In this threat model, the attacker can poison a small fraction of the training data to embed a secret functionality in the trained model. Even though our experiments use two different versions of a given neural network (one version with GAP and one without), we do not assume that the attacker controls the network architecture. We want to 1) explore whether a GAP layer makes the adversary stronger by allowing dynamic backdoor attacks and 2) raise awareness about this ``vulnerability'' to the community.
\section{Experimental Results}
\label{sec:results}

We run experiments that use a trigger in different positions between training and inference using the networks described in~\autoref{ssec:nn}. Every experiment was run ten times to limit the effects of stochastic gradient descent's randomness.

For a fair comparison of the results, we have to make sure that, in every case, both versions of the network perform similarly. We trained all the models with a clean dataset and show the results in the second rightmost column of~\cref{tab:sound-clean-acc,tab:text-clean-acc,tab:image-clean-acc}. In most cases, GAP slightly improves the model's performance ($\sim1$\%), meaning that GAP's feature map averaging improves the model's generalization. Only in one case (small CNN in sound recognition and 30 classes), the model's performance is dropped significantly ($\sim3$\%). In this case, the network's capacity is greatly reduced by the GAP layer, meaning that the network cannot encode effectively all the information included in the entire speech commands dataset.

An effective backdoor should remain hidden while the network operates on clean inputs because, in any other case, the user could become suspicious and stop using the backdoored model. As a result, the adversary needs to ensure no significant performance drop for clean inputs from the backdoor insertion. The two rightmost columns in~\cref{tab:sound-clean-acc,tab:text-clean-acc,tab:image-clean-acc} show the accuracy of clean and poisoned models for clean inputs. By comparing the rows of these two columns in~\cref{tab:sound-clean-acc,tab:text-clean-acc,tab:image-clean-acc}, we see that in most cases, the performance drop is less than 1\%, which can remain unnoticed.

In~\cref{fig:attack-acc}, we plot the results for our experiments. In each of these graphs, the straight lines represent the network with GAP and the dotted ones the network without it.

%

\begin{table}[ht!]
    \centering
    \caption{Clean accuracy drop in sound recognition.}
    \label{tab:sound-clean-acc}
    \resizebox{.30\textwidth}{!}{%
        \begin{tabular}{|c|c|c|c|c|}
        \hline
        CNN & classes & type & original & poisoned \\ \hline
        \multirow{4}{*}{large} & \multirow{2}{*}{10} & FC & 95.56 ($\pm$ 0.172) & 94.92 ($\pm$ 0.368) \\ \cline{3-5} 
        & & GAP &  96.18 ($\pm$ 0.22) & 95.81 ($\pm$ 0.31)  \\ \cline{2-5}
        & \multirow{2}{*}{30} & FC & 94.80 ($\pm$ 0.288) & 94.64 ($\pm$ 0.437)  \\ \cline{3-5} 
        & & GAP & 95.60 ($\pm$ 0.09) & 95.67 ($\pm$ 0.21)  \\ \hhline{|=|=|=|=|=|}
        \multirow{4}{*}{small} & \multirow{2}{*}{10} & FC & 90.27 ($\pm$ 0.369) & 89.56 ($\pm$ 0.458) \\ \cline{3-5} 
        & & GAP & 91.40 ($\pm$ 0.41) & 90.27 ($\pm$ 0.80) \\ \cline{2-5} 
        & \multirow{2}{*}{30} & FC & 88.06 ($\pm$ 0.451) & 87.44 ($\pm$ 0.376)  \\ \cline{3-5} 
        & & GAP &  85.75 ($\pm$ 0.63) & 86.15 ($\pm$ 0.58) \\ \hline
        \end{tabular}%
    }
\end{table}

\begin{table}[]
    \centering
    \caption{Clean accuracy drop for text sentiment analysis.}
    \label{tab:text-clean-acc}
    \resizebox{.30\textwidth}{!}{%
        \begin{tabular}{|c|c|c|c|}
        \hline
        architecture & type & original & poisoned \\ \hline
        \multirow{2}{*}{$1^{st}$} & FC & 84.64 ($\pm$ 0.398) & 84.86 ($\pm$ 0.424) \\ \cline{2-4} 
        & GAP &  84.01 ($\pm$ 0.41) & 83.88 ($\pm$ 0.45) \\ \hhline{|=|=|=|=|}
        \multirow{2}{*}{$2^{nd}$} & FC & 86.14 ($\pm$ 0.698) & 85.96 ($\pm$ 0.811)\\ \cline{2-4} 
         & GAP &  86.56 ($\pm$ 0.21) & 86.46 ($\pm$ 0.38)  \\ \hhline{|=|=|=|=|}
        \multirow{2}{*}{$3^{rd}$} & FC & 85.60 ($\pm$ 0.366) & 85.54 ($\pm$ 0.325) \\ \cline{2-4} 
         & GAP & 86.13 ($\pm$ 0.06) & 86.13 ($\pm$ 0.08)\\ \hline
        \end{tabular}%
    }
\end{table}

\begin{table}[]
    \centering
    \caption{Clean accuracy drop for image classification.}
    \label{tab:image-clean-acc}
    \resizebox{.30\textwidth}{!}{%
        \begin{tabular}{|c|c|c|c|}
        \hline
        architecture & type & original & poisoned \\ \hline
        \multirow{2}{*}{STRIP} & FC & 86.26 ($\pm$ 0.419) & 86.16 ($\pm$ 0.315) \\ \cline{2-4} 
        & GAP &  87.26 ($\pm$ 0.271) & 87.25 ($\pm$ 0.277) \\ \hline
        \end{tabular}%
    }
\end{table}

\subsection{Sound Recognition}

\begin{figure*}[t]
    \centering
    \subfloat[][Large CNN (10 classes)]{    \includegraphics[width=0.29\textwidth]{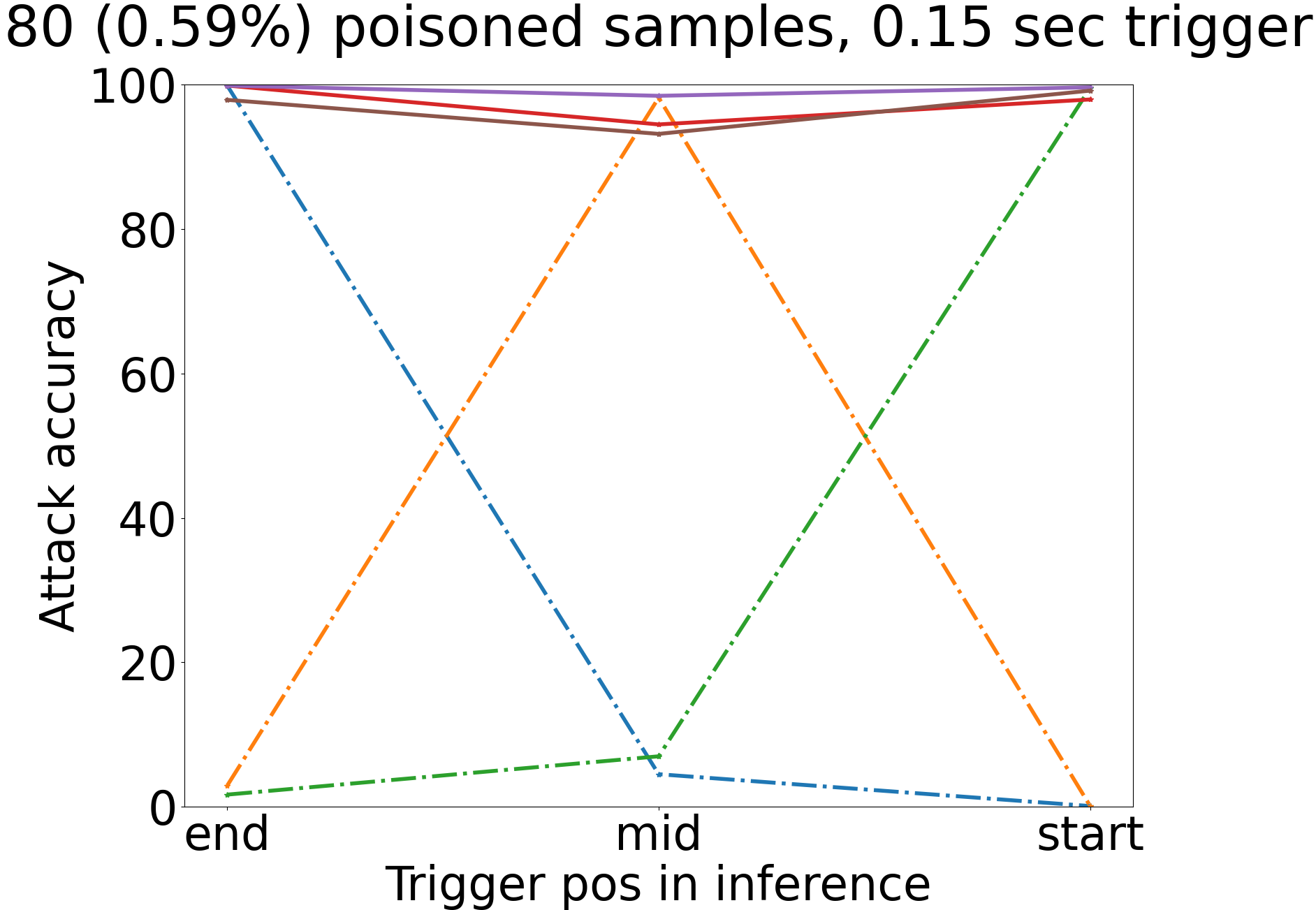}%
    \label{subfig:sound-attack-acc-large-cnn-10}}
    \hfil
    \subfloat[][Large CNN (30 classes)]{    \includegraphics[width=0.29\textwidth]{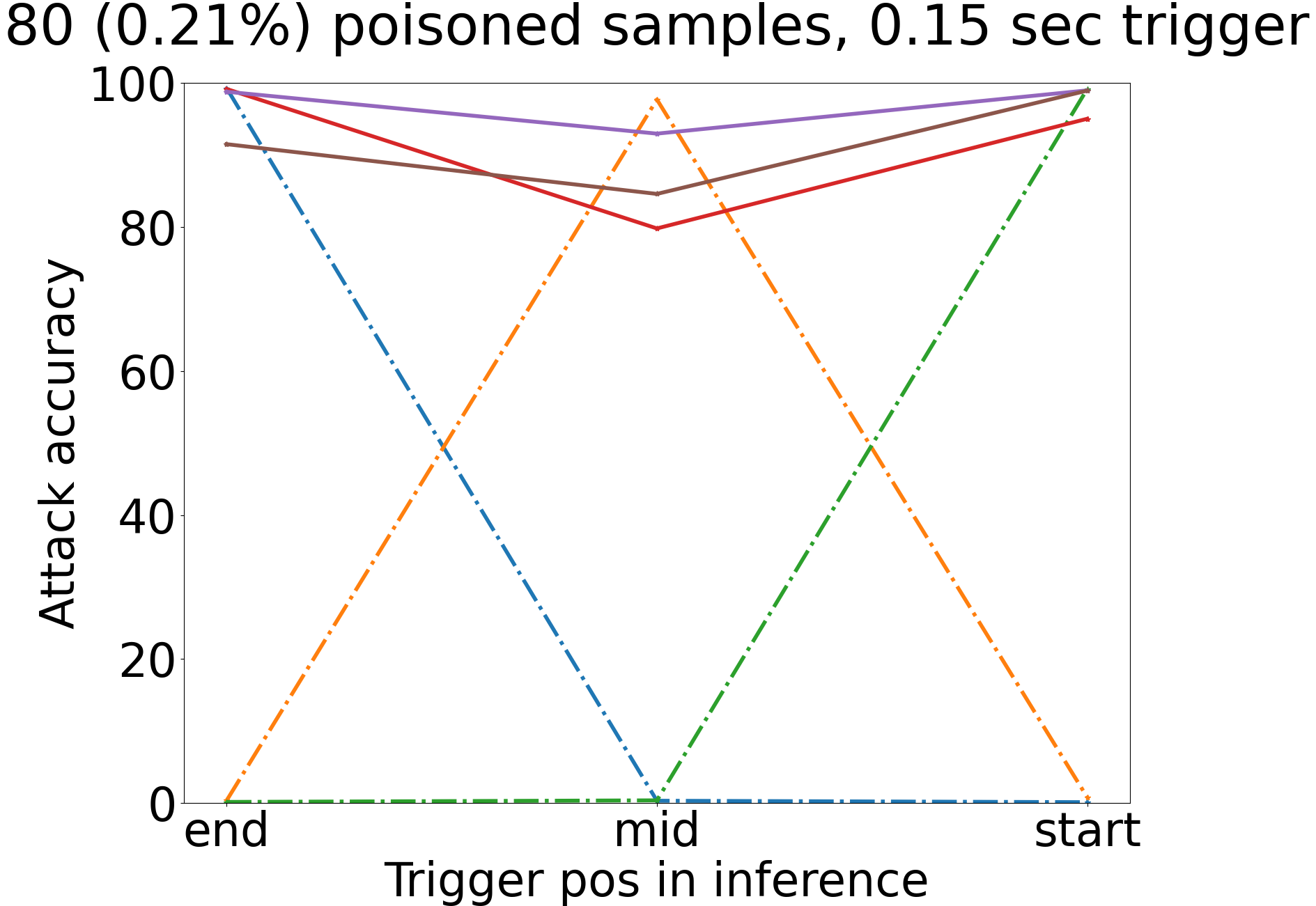}%
    \label{subfig:sound-attack-acc-large-cnn-30}}
    \hfil
    \subfloat[][Small CNN (10 classes)]{    \includegraphics[width=0.29\textwidth]{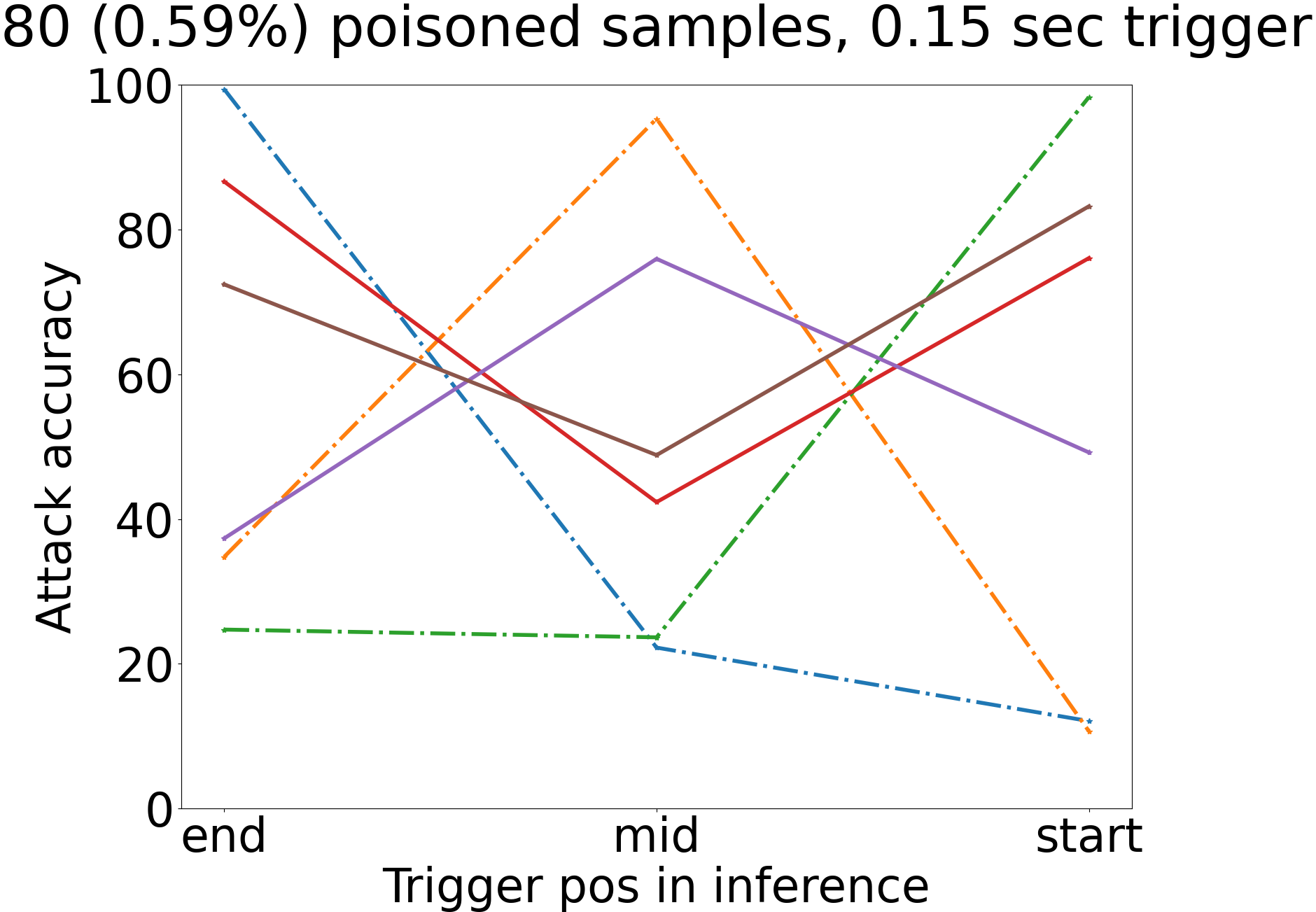}%
    \label{subfig:sound-attack-acc-small-cnn-10}}
    \hfil \\
    \subfloat[][First architecture in text]{
    \includegraphics[width=0.29\textwidth]{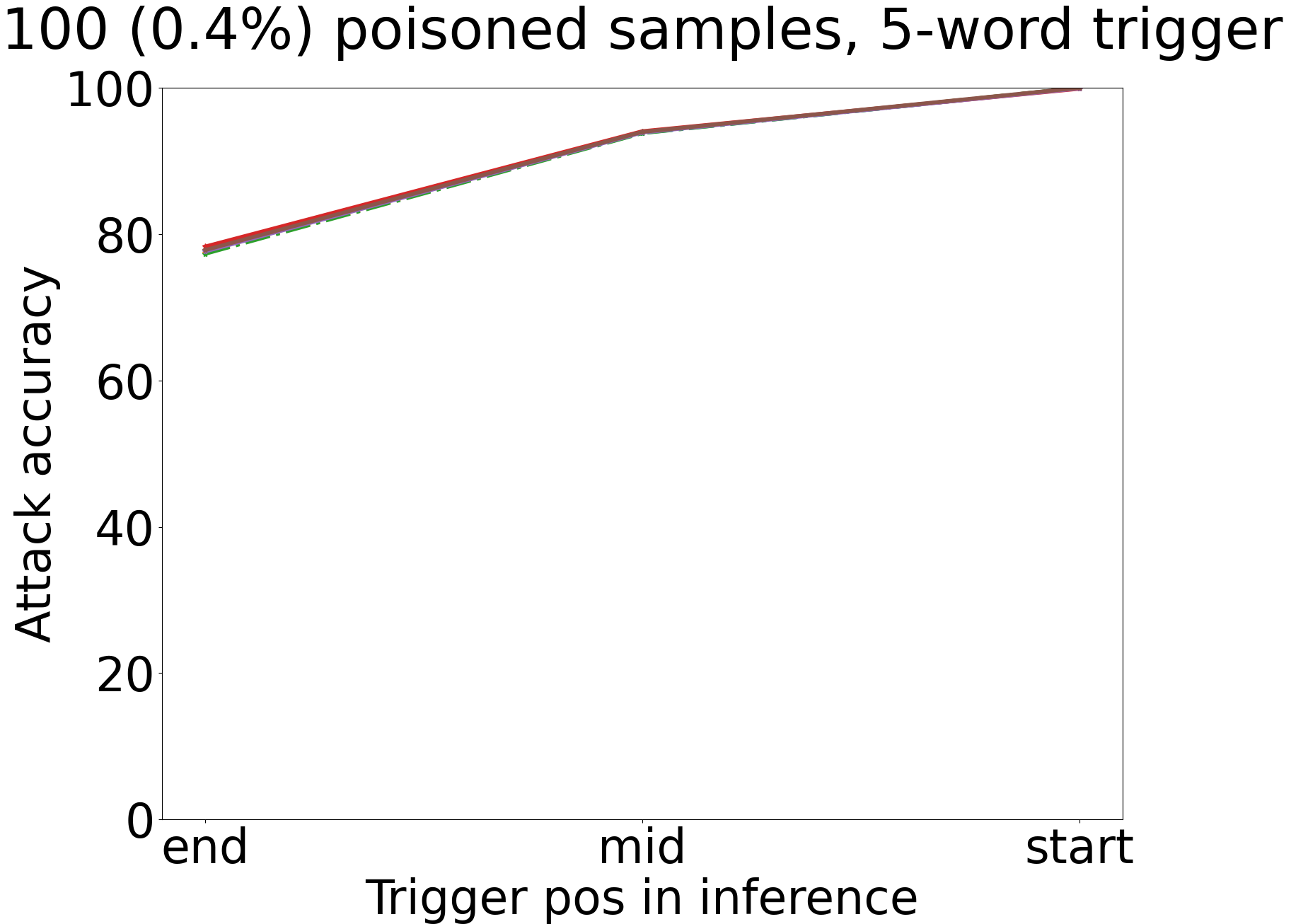}%
    \label{subfig:txt-attack-acc-troj}}
    \hfil
    \subfloat[][Second architecture in text]{    \includegraphics[width=0.39\textwidth]{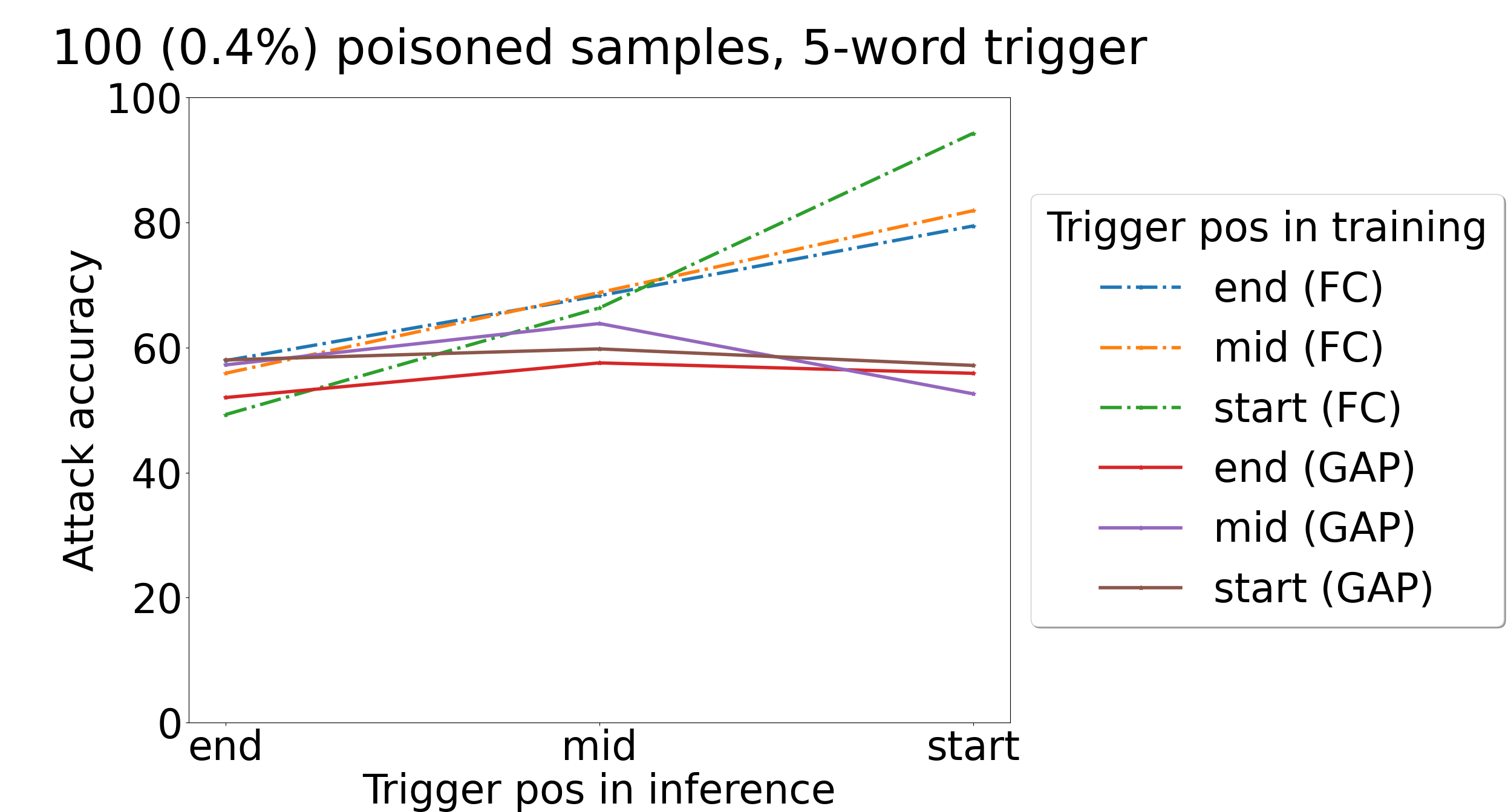}%
    \label{subfig:txt-attack-acc-github}}
    \hfil
    \caption{Attack accuracy for different triggers between training and inference in sound and text classification.}
    \label{fig:attack-acc}
\end{figure*}

\Cref{subfig:sound-attack-acc-large-cnn-10,subfig:sound-attack-acc-large-cnn-30} show the attack accuracy for different triggers in sound recognition for the large CNN.
In both cases, we see that when GAP is omitted, the backdoor attack is successful only when the same trigger is used for training and inference. However, when GAP is used, the position of the trigger becomes unimportant, and the backdoor attack is successful for all the triggers we tried. In this architecture, the network's capacity remains high even with a GAP layer, meaning that most of the dataset's information can still be encoded in its weights. Additionally, GAP's feature map averaging makes the learned representation spatially invariant. Thus, an adversary could create a stronger backdoor attack under the same threat model without requiring access to more data just by exploiting GAP's properties.

%

%

On the other hand, we saw a different behavior for the small CNN. GAP makes the attack impossible when the entire dataset is used, even if the same trigger is used in training and
testing. In this case, the model's capacity is very small, and it cannot learn useful information from only a few poisoned samples. Then, only an increase in the poisoning rate could lead to an increase in the attack accuracy. When we use ten classes (\cref{subfig:sound-attack-acc-small-cnn-10}), GAP introduces some spatial invariance in the trigger as the attack is, in general, more successful when different triggers are used in training and inference. Still, GAP results in a lower attack success rate when the same trigger is used in training and testing because its generalization prevents any overfitting required for a successful backdoor attack.



\subsection{Text Sentiment Analysis}

For the first architecture (\cref{subfig:txt-attack-acc-troj}), the attack performs identically for both versions tried. As we discussed in~\cref{ssec:nn}, the two network versions are very similar, and the GAP layer does not introduce any noticeable differences. Additionally, in~\cref{subfig:txt-attack-acc-troj}, the trigger's position is unimportant for both versions. This architecture uses max-over-time-pooling and finds the most important phrases of 3, 4, or 5 words in a sentence. Thus, in most cases, our 5-word trigger can be spotted easily.  


The two remaining architectures in text sentiment analysis show similar behavior, and for that reason, we plot only the results (\cref{subfig:txt-attack-acc-github}) for the second architecture (\cref{tab:nn-txt-mata}). First, GAP makes the trigger almost equally effective even if it is injected in different positions in training and testing. However, GAP's generalization makes the attack more difficult in general, and the attack success rate does not surpass 65\% in our experiments. This percentage is increased only by poisoning more data or allowing overfitting. For example, if we continue training even if validation loss increases, the attack success rate increases up to 85\%. This indicates that a backdoor could be harder for models affected by an average of their inputs and not only by specific features.
On the other hand, the trigger's position can play a crucial role in the attack success rate when GAP is not used. The network can easily spot the trigger when inserted at the sentence's beginning in the test time, regardless of its position in training. Still, the attack success rate is lower in any other case.



%

\subsection{Image Classification}

To test our hypothesis in image classification, we poisoned 100 samples (0.2\%). We saw that GAP alone could not create a dynamic backdoor attack as the attack accuracy is high ($\geq90\%$) only for the same triggers in training and testing. It slightly increases the attack accuracy in three cases at 33\%, but the introduced spatial invariance is not strong enough for a successful dynamic backdoor attack. This number is not increased notably even if we significantly increase the poisoning rate. Concluding, in deeper CNNs, it is very challenging to create dynamic backdoors with a GAP layer as they have already discarded spatial information through the convolution filters.
\section{Conclusions and Future Work}
\label{sec:conclusions}

To the best of our knowledge, we are the first to experiment with different triggers in training and inference in sound, text, and image classification. These experiments led to various useful insights. First, we showed that a GAP layer could lead to dynamic backdoor attacks in neural networks. This is a very rare event, though, as the network's capacity should remain large enough to encode the dataset's information but low enough to avoid learning particular data relations. 

Furthermore, we saw that the backdoor attack could be effective when the model can overfit the data and ineffective when the model's generalization is strong. Thus, robust generalization techniques and methods that prevent learning from features in very few training samples could result in effective defenses, which is an interesting future research direction. Additionally, we plan to exploit GAP's properties to bypass existing defenses based on the static nature of the triggers and the feature maps of the last network layers.


\printbibliography

\end{document}